\documentclass[aps,floats]{revtex4}
\usepackage{amsmath,amssymb}
\usepackage{graphicx,epsfig}

\begin{document}
\bibliographystyle {plain}

\def\oppropto{\mathop{\propto}} 
\def\opsimeq{\mathop{\simeq}}
\def\opoverderline{\mathop{\overline}}
\def\operarrow{\mathop{\longrightarrow}}
\def\opsim{\mathop{\sim}}

\def\fig#1#2{\includegraphics[height=#1]{#2}}
\def\figx#1#2{\includegraphics[width=#1]{#2}}


\title{ Random walk in two-dimensional self-affine random potentials : \\
 strong disorder renormalization approach } 


 \author{ C\'ecile Monthus and Thomas Garel }
  \affiliation{ Institut de Physique Th\'{e}orique, CNRS and CEA Saclay,
 91191 Gif-sur-Yvette, France}

\begin{abstract}
We consider the continuous-time random walk of a particle in a two-dimensional self-affine quenched random potential of Hurst exponent $H>0$. The corresponding master equation is studied via the strong disorder renormalization procedure introduced in  Ref. [C. Monthus and T. Garel, J. Phys. A: Math. Theor. 41 (2008) 255002]. We present numerical results on the statistics of the equilibrium time $t_{eq}$ over the disordered samples of a given size $L \times L$ for $10 \leq L \leq 80$. We find an 'Infinite disorder fixed point', where the equilibrium barrier $\Gamma_{eq} \equiv \ln t_{eq}$ scales as $\Gamma_{eq}=L^H u $ where $u$ is a random variable of order $O(1)$. This corresponds to a logarithmically-slow diffusion $ \vert \vec r(t) - \vec r(0) \vert \sim (\ln t)^{1/H}$ for the position $\vec r(t)$ of the particle.
\end{abstract}

\maketitle

\section{ Introduction} 

 Random walks and diffusion processes have been the subject of
 constant interest in mathematics and in physics
during the last century,  for two main reasons 
 (i) they play a central role in probability theory,
and present a large number of very nice mathematical properties
(ii)  they naturally appear
 in a great variety of situations in physics and in biology.
It is thus important to understand  
the effects of quenched disorder on random walks :
 are the usual properties of random walks stable with respect
 to the presence of some disorder or inhomogeneity ?
 If not, what are the new properties induced by disorder?
Among the various types of random walks in random media 
that have been considered in the past (see the reviews \cite{HK,HBA,jpbreview} 
and references therein), we wish to focus here on the case
of random walks in a two-dimensional self-affine random potential $U(\vec r)$.
In a continuous framework, this model can be defined via the Langevin equation
for the position $\vec r$ of the particle
\begin{eqnarray}
\frac{d \vec r}{dt} =  - \vec \nabla U(\vec r) + \vec \eta (t)
\label{Langevin}
\end{eqnarray}
where $\vec \eta(t) $ is a white noise 
\begin{eqnarray} 
< \eta_i(t) \eta_j(t ') > = 2 T \delta(t-t ') \delta_{i,j}
\end{eqnarray}
that would generate a Brownian diffusion
in the absence of the random potential $U$, and where the quenched random potential
$U(\vec r)$ is self-affine with some Hurst exponent $H$
\begin{eqnarray}
\overline{ \left[ U(\vec r) -U(\vec r \ ') \right]^2 }
\opsimeq_{ \vert \vec r - \vec r \ ' \vert \to \infty}
 \vert \vec r - \vec r \ ' \vert^{2H}
\label{correU2d}
\end{eqnarray}
The case of dimension $d=1$ and Hurst exponent $H=1/2$ 
corresponds to the random-force Sinai model where the logarithmically-slow behavior
$\vert r(t) -r(0) \vert \sim (\ln t)^{2}$ has been obtained
 via various exact methods 
(see for instance the review \cite{jpbreview} and references therein).
Since this logarithmic behavior replaces 
the usual power-law behaviour  $x \sim \sqrt{t}$ of the pure Brownian motion,
the effect of disorder is extremely strong.
In higher dimension $d>1$, the model is not exactly solvable,
but from scaling arguments on barriers,
one still expects the analogous
 logarithmic scaling \cite{Mar83,jpbreview}
\begin{eqnarray}
 \vert \vec r(t) - \vec r(0) \vert \sim (\ln t)^{1/H}
\label{logslow}
\end{eqnarray}
However, to the best of our knowledge, this behavior has not been much tested,
 except in the preliminary unpublished numerical results of Pettini
shown on Fig. 4.9 of the review \cite{jpbreview}.
The aim of this paper is to study a continuous-time lattice version
of this model in dimension $d=2$, via 
the strong disorder renormalization procedure introduced in \cite{rgmaster}
that can be applied to any master equation in arbitrary dimension.

The paper is organized as follows.
In section \ref{sec-wm}, we recall the Weierstrass-Mandelbrot function method
to generate numerically two-dimensional self-affine random potentials.
In section \ref{sec-rg}, we explain how to use for the present case
the strong disorder renormalization method introduced in \cite{rgmaster}.
In section \ref{sec-nume}, we present our numerical results concerning
the statistics of the equilibrium time $t_{eq}$
over the disordered samples of a given size $L \times L$.
Our conclusions are summarized in section \ref{sec-conclusion}.

\section{ Method for generating a two-dimensional
 self-affine random potential   }

\label{sec-wm}

Among the various methods
that have been proposed in the literature to generate
random functions of a given Hurst exponent 
 (see the reviews \cite{voss_saupe}
and a comparative study of their performances in \cite{jennane}),
we have found numerically that the method giving the best results for the correlation
of Eq. \ref{correU2d}
is the so-called Weierstrass-Mandelbrot function method, that we recall in this section.

\begin{figure}[htbp]
 \includegraphics[height=6cm]{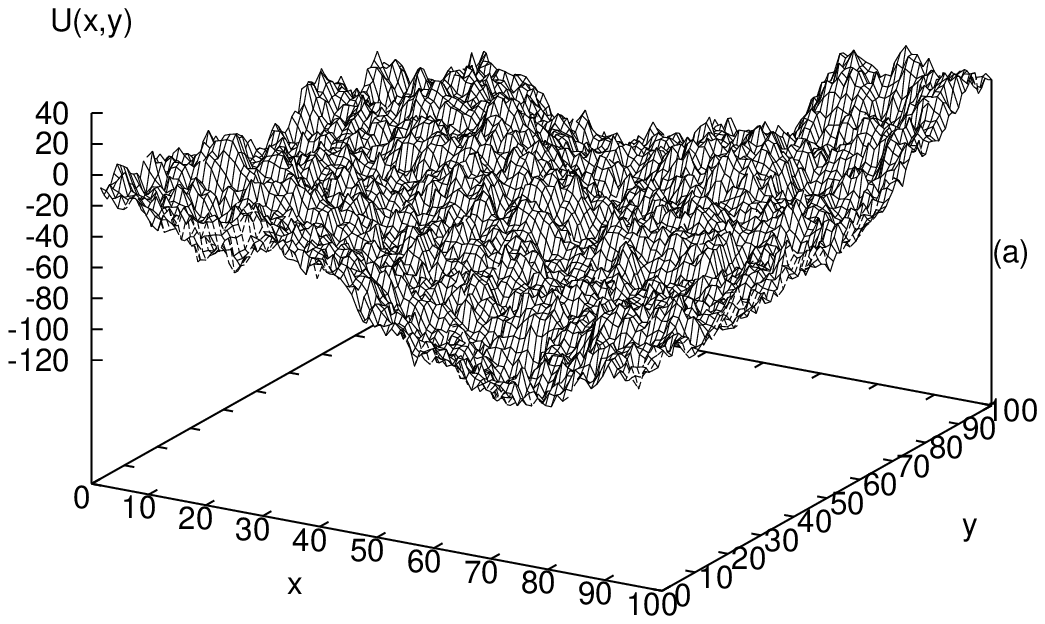}
\vspace{1cm}
 \includegraphics[height=6cm]{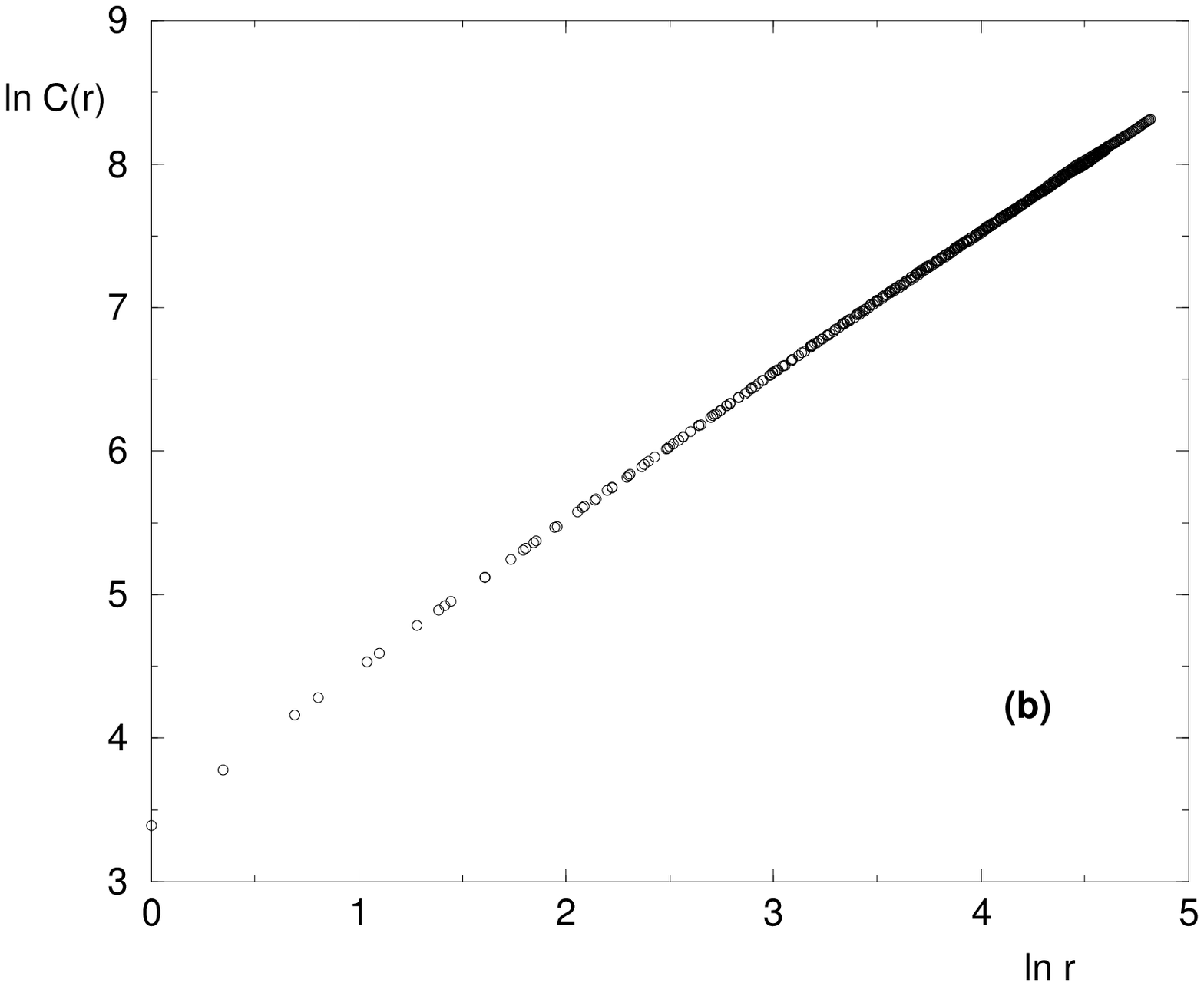}
\caption{ Random self-affine random potential $U(x,y)$
on a square of size $100 \times 100$,
obtained via the method of Eq. \ref{wm2d} for the value $H=0.5$ of the Hurst exponent :
(a) Example of one realization of the random potential  $U(x,y)$ .
(b) Log-log plot of the correlation function $C(r) \equiv \overline{ \left[ U(\vec r_1) -U(\vec r_2) \right]^2} $ as a function of the distance $r \equiv \vert \vec r_1 -\vec r_2 \vert$ after averaging over angles and over disorder realizations : the slope is here $2H=1$, as it should for $H=0.5$ (see Eq. \ref{correU2d}). }
\label{figpotential}
\end{figure}

\subsection{ Reminder on the Weierstrass-Mandelbrot function in dimension $d=1$ }

In dimension $d=1$, the Weierstrass-Mandelbrot function is defined by
\cite{mandelbrot,berry}
\begin{eqnarray}
U(x) = \sum_{n=n_{min}}^{n_{max}}  \frac{ \cos(2 \pi \phi_n)-\cos \left( 2 \pi \gamma^n x + 2 \pi \phi_n \right) }{\gamma^{n H} }
\label{wm1d}
\end{eqnarray}
where the phases $\phi_n$ are independent and uniform in $[0,1]$,
and where $n_{min}=-\infty$ and $n_{max}=+\infty$. The function $U(x)$ is fractal 
with Hurst exponent $H$ on all scales : the frequencies $\gamma^n$
are in geometric progression, in contrast with a Fourier transform
that would correspond to an arithmetic progression. 
In the limit $\gamma \to 1$, the discrete spectrum become dense
and the function $U(x)$ converge towards the fractional Brownian motion
of exponent $H$. We refer to \cite{berry} for more details on its mathematical properties
and now discuss how to use it for numerical simulations.

If one wishes to generate the potential $U(x)$ at $N$ discrete points 
 $x=1,2,...N$,  one has to choose the three parameters $(n_{min},n_{max},\gamma)$
in the following way :

(i) the maximal Fourier frequency $\omega_{max}$ associated to the lattice
spacing $\Delta x=1$ is $\omega_{max}=1/(\Delta x)=1$. A convenient choice
is thus $n_{max}=0$, corresponding to the maximal frequency $\gamma^{n_{max}} =1 $
in the sum of Eq. \ref{wm1d}.

(ii) the minimal Fourier frequency $\omega_{min}$ associated to the sample size
$N$  is $\omega_{min}=1/N$. Since we do not wish any periodicity of order $N$
in the potential, we have to choose $n_{min}$ such that the minimal frequency 
$\gamma^{n_{min}}$ in the sum of Eq. \ref{wm1d} satisfies $\gamma^{n_{min}} \ll 1/N  $.

(iii) finally, the parameter $\gamma$ determines the discretization
of the frequency spectrum : the frequency $\gamma^n$ have to be sufficiently dense.

We now turn to the generalization to higher dimension.

\subsection{ Generalization to dimension $d=2$ }

To generalise Eq. \ref{wm1d} to higher dimension $d>1$,
the idea \cite{ausloos,yan} is to keep the principle 
of a sum over plane waves of various vectors $\vec k$,
where the modulus $\vert \vec k \vert$ varies in geometric progression $\gamma^n$,
and where the angular part of $\vec k$ is uniformly distributed to insure isotropy.
In dimension $d=2$, this corresponds to \cite{ausloos,yan}
\begin{eqnarray}
U(x,y) = \sum_{n=n_{min}}^{n_{max}} \frac{1}{\sqrt m_{max}} \sum_{m=1}^{m_{max}}
\frac{ \cos(2 \pi \phi_{n,m}) -\cos \left( 2 \pi \gamma^n (x \cos 2 \pi\alpha_{n,m}+y \sin 2 \pi\alpha_{n,m}) + 2 \pi \phi_{n,m} \right) }{\gamma^{n H} }
\label{wm2d}
\end{eqnarray}
where the phases $\alpha_{n,m}$
and $\phi_{n,m}$ are independent and uniformly distributed in $[0,1]$.
The new parameter $m_{max}$ fixes the number of wave vectors $\vec k$
of a given modulus $\vert \vec k \vert$.
We have checked that this generalization proposed in 
\cite{ausloos,yan} gives satisfactory numerical realizations
of self-affine random potential
(whereas the alternative generalization proposed in \cite{lopez,jennane2D}
that are based on cartesian coordinates presents anisotropy).

Here we wish to generate the potential $U(x,y)$ at $N^2$ discrete points 
where $x=1,2,...N$ and $y=1,2,..,N$. 
One has then to choose the four parameters $(n_{min},n_{max},m_{max},\gamma)$
to obtain good results for the two-point function of Eq. \ref{correU2d}
for all pairs of points of the samples.
For squares samples of linear size $10 \leq L \leq 80$, we have found that the following
set of parameters give satisfactory realizations of the potential $U(x,y)$ 
for Hurst exponents $0.3 \leq H \leq 0.8$ :
$n_{max}=0$, $n_{min}=-150$, $m_{max}=100$, $\gamma=1.2$.
We show on Fig. \ref{figpotential} (a) 
an example of realization of the random self-affine random potential $U(x,y)$
on a square of size $100 \times 100$,
for the value $H=0.5$ of the Hurst exponent.
The corresponding correlation function is shown on Fig. \ref{figpotential} (b) 
on a log-log plot.

\section{ Strong disorder renormalization procedure }

\label{sec-rg}

Strong disorder renormalization 
(see \cite{review} for a review) is a very specific type of RG
that has been first developed in the field of quantum spins :
the RG rules of Ma and Dasgupta \cite{madasgupta} 
have been put on a firm ground by D.S. Fisher 
who introduced the crucial idea of ``infinite disorder'' fixed point
where the method becomes asymptotically exact,
and who computed explicitly exact 
critical exponents and scaling functions 
for one-dimensional disordered quantum spin chains \cite{dsf}.
This method has thus generated a lot of activity for various
disordered quantum models \cite{review}, and has been then
successfully applied to
various classical disordered dynamical models,
such as random walks in random media
 \cite{sinairg,sinaibiasdirectedtraprg},
reaction-diffusion in a random medium \cite{readiffrg}, 
coarsening dynamics of classical spin chains \cite{rfimrg}, 
trap models \cite{traprg}, random vibrational networks \cite{vibrational},
absorbing state phase transitions \cite{contactrg},
zero range processes \cite{zerorangerg} and 
exclusion processes  \cite{exclusionrg}.

For random walks in random media, the procedure introduced in Refs 
\cite{sinairg,sinaibiasdirectedtraprg} 
or in the recent work \cite{Jack_Soll} are specific to the dimension $d=1$.
Here in dimension $d=2$, the appropriate framework 
is the 'strong disorder renormalization' 
(RG) procedure introduced \cite{rgmaster} 
 that can be defined for any master equation.
In this section, we recall its principles
 for the present problem of a particle in a two-dimensional potential.

\subsection{ Master Equation}

The master equation describing the evolution of the
probability $P_t ({\vec r} ) $ to be at  position ${\vec r}$
 at time t can be written as
\begin{eqnarray}
\frac{ dP_t \left({\vec r} \right) }{dt}
= \sum_{\vec r \ '} P_t \left({\vec r}\ ' \right) 
W \left({\vec r}\ ' \to  {\vec r}  \right) 
 -  P_t \left({\vec r} \right) W_{out} \left( {\vec r} \right)
\label{master}
\end{eqnarray}
where $ W \left({\vec r}\ ' \to  {\vec r}  \right) $ 
represents the transition rate per unit time from position 
${\vec r}\ '$ to ${\vec r}$, and 
\begin{eqnarray}
W_{out} \left( {\vec r} \right)  \equiv
 \sum_{ {\vec r} '} W \left({\vec r} \to  {\vec r}\ ' \right) 
\label{wcout}
\end{eqnarray}
represents the total exit rate out of position ${\vec r}$.

For the two-dimensional random walk in the random potential $U(\vec r)$
at temperature $T$, we have chosen to consider 
the Metropolis dynamics
defined by the transition rates 
\begin{eqnarray}
W \left( \vec r \to \vec  r \ '  \right)
= \delta_{<\vec r, \vec r\ ' >} 
\  {\rm min} \left(1, e^{-  (U(\vec r \ ' )-U(\vec r ))/T } \right)
\label{metropolis}
\end{eqnarray}
The first factor $\delta_{<\vec r, \vec r\ ' >}$
 means that the two positions
are neighbors on the two-dimensional lattice, and
 the last factor ensures the convergence towards thermal equilibrium
at temperature $T$ via the detailed balance property
\begin{eqnarray}
 e^{- U(\vec r) /T} W \left( \vec r \to \vec r\ '  \right)
= e^{- U(\vec r\ ') /T} W \left( \vec r\ ' \to \vec r  \right)
\label{detailedbalance}
\end{eqnarray}

\subsection{ Strong disorder renormalization rules }

For dynamical models, the aim of any renormalization procedure
is to integrate over 'fast' processes to obtain effective properties 
of 'slow' processes.
 The general idea of 'strong renormalization' for dynamical models
consists in eliminating iteratively the 'fastest' process.
The RG procedure introduced
in \cite{rgmaster} can be summarized as follows :

(1) find the position $\vec r^*$ with the largest exit rate $W^*_{out}$
\begin{eqnarray}
W^*_{out} = W_{out} \left( \vec r^* \right)
 \equiv  {\rm max}_{\vec r} \left[  W_{out} \left( \vec r \right) \right]
\label{defwmax}
\end{eqnarray}

(2) find the neighbors $(\vec r_1,\vec r_2,...,\vec r_n)$
 of position $\vec r^*$, 
i.e. the surviving positions that are related 
via positive rates $W(\vec r^* \to \vec r_i )>0$ and
 $W(\vec r_i\to \vec r^*)>0$
to the decimated position $\vec r^*$.
For each neighbor position $\vec r_i$ with
$i \in (1,..,n)$, update the transition rate to go to
the position $\vec r_j$ with $j \in (1,..,n)$ and $j \neq i$
 according to
\begin{eqnarray}
W^{new}(\vec r_i \to \vec r_j )=W(\vec r_i \to \vec r_j )
+ W(\vec r_i \to \vec r^* ) \times  \pi_{\vec r^*} \left(\vec r_j \right) 
\label{wijnew}
\end{eqnarray}
where the first term represents the 'old' transition rate (possibly zero),
and the second term represents the transition 
via the decimated position $\vec r^*$ :
the factor $W(\vec r_i \to \vec r^* ) $ takes into account 
the transition rate to $\vec r^*$ and the term
\begin{eqnarray}
\pi_{\vec r^*} \left(\vec r_j \right) = 
\frac{W \left(\vec r^*  \to \vec r_j \right)}{W_{out} \left( \vec r^*\right)}
\label{picstar}
\end{eqnarray}
represents the probability
 to make a transition towards $\vec r_j$
when in $\vec r^*$.
The $2 n$ rates $W(\vec r^* \to \vec r_i )$
 and $W(\vec r_i \to \vec r^*)$ then
 disappear with the decimated position $\vec r^*$.
Note that the rule of Eq. \ref{wijnew} 
has been recently proposed in \cite{vulpiani}
to eliminate 'fast states'  from various dynamical problems 
with two very separated time scales.
The physical interpretation of this rule is as follows :
the time spent in the decimated position $\vec r^*$ is neglected
with respects to the other time scales remaining in the system. 
The validity of this approximation within the present renormalization procedure
is discussed in detail in \cite{rgmaster}. 

(3) update the exit rates out of the neighbors $\vec r_i$ of $\vec r^*$, with $i=1,..,n$
either with
the definition
\begin{eqnarray}
W^{new}_{out}(\vec r_i)   = 
\sum_{\vec r} W^{new}(\vec r_i \to \vec r )
\label{wioutnewactualisation}
\end{eqnarray}
or with the rule that can be deduced from Eq. \ref{wijnew}
 \begin{eqnarray}
W^{new}_{out}(\vec r_i) = W_{out}(\vec r_i) 
 - W(\vec r_i \to \vec r^* ) 
\frac{ W(\vec r^* \to \vec r_i )}
{W^{*}_{out}}
\label{wioutnew}
\end{eqnarray}
The physical meaning of this rule is the following.
The exit rate out of the position $\vec r_i$ decays because 
the previous transition towards $\vec r^*$ can lead to an immediate return
towards $\vec r_i$. After the decimation of the position $\vec r^*$,
this process is not considered as an 'exit' process anymore, but as a
residence process in the position $\vec r_i$.
This point is very important to understand the
 meaning of the renormalization procedure :
the remaining positions at a given renormalization scale are 
'formally' microscopic positions 
of the initial master equation (Eq. \ref{master}),
but each of these remaining microscopic position
 actually represents some 'valley' in position space
that takes into account all the previously decimated positions.

(4) return to point (1).

We refer to \cite{rgmaster} for more detailed explanations.
In practice, the renormalized rates $W(\vec r \to \vec r \ ' )$
can rapidly  become very small as a consequence of the multiplicative structure
of the renormalization rule of Eq \ref{wijnew}. This means that the appropriate
 variables are the logarithms of the transition rates, that we will call 'barriers' in the remaining
of this paper. The barrier $B (\vec r \to \vec r \ ' )$ from $\vec r$  to $\vec r\ ' $ is defined by
\begin{eqnarray}
B (\vec r \to \vec r\ ' )\equiv - \ln W(\vec r \to \vec r\ ' )
\label{defb}
\end{eqnarray}
and similarly the exit barrier out of position $\vec r$ is defined by
\begin{eqnarray}
B_{out} (\vec r )\equiv - \ln W_{out}(\vec r )
\label{defbout}
\end{eqnarray}
A very important advantage of this formulation in terms
of the renormalized transition rates of the master equation is that 
the renormalized barriers take into account the true 'barriers'
of the dynamics, whatever their origin which can be
 either energetic or entropic.

\subsection{ Numerical details }

We have applied numerically these renormalization rules
for square samples of size $L^2$ with $10 \leq L \leq 80$
with a statistics of  $33.10^5 \geq n_s(L) \geq 280$  disordered samples.
We have studied six values of the Hurst exponent in the interval $0.3 \leq H \leq 0.8$.

\section{ Statistics of the equilibrium time of finite systems }

\label{sec-nume}

\begin{figure}[htbp]
 \includegraphics[height=6cm]{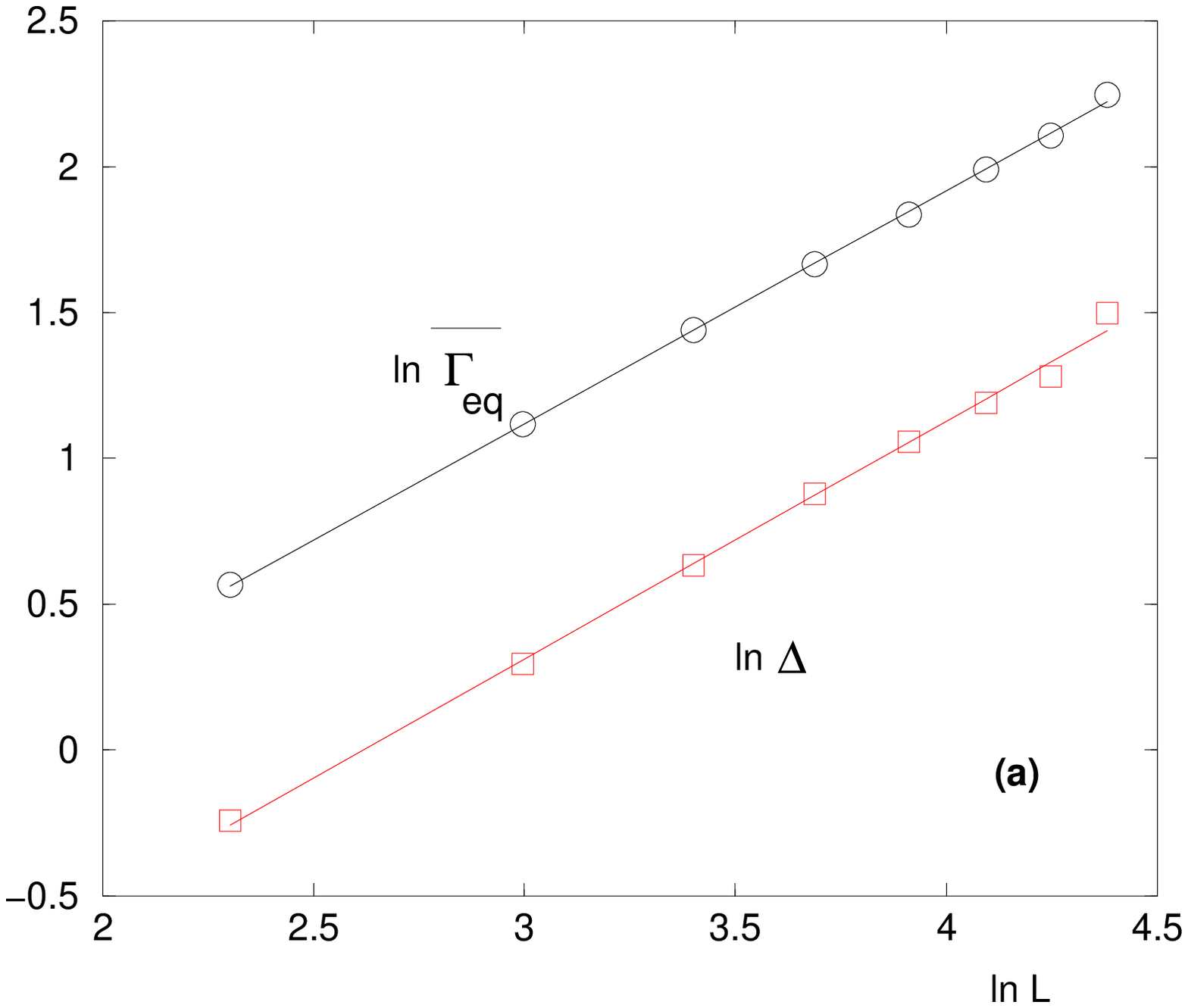}
\hspace{2cm}
 \includegraphics[height=6cm]{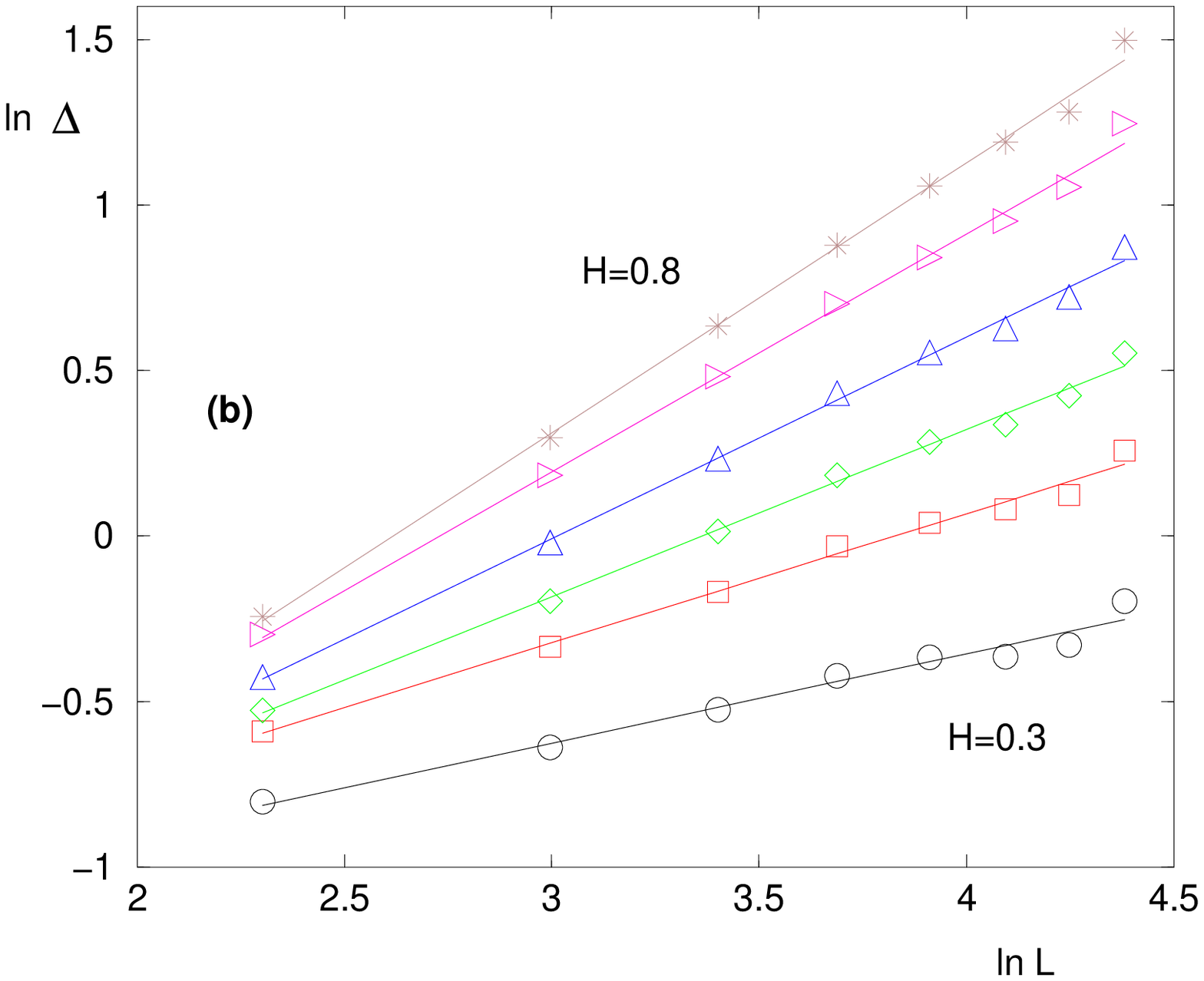}
\vspace{1cm}
\caption{ (Color on line) 
Statistics of the equilibrium time $t_{eq}$ over the disordered samples of sizes $L^2$ : the appropriate variable is $\Gamma_{eq}= \ln t_{eq}$ (Eq. \ref{defgammaeq})
(a) the disorder-averaged value $\overline{\Gamma_{eq}}(L)$
and the width $\Delta(L)$ shown here for $H=0.8$ scale with the same exponent
$\psi$ (see Eq. \ref{psiexp}).
(b) the exponent $\psi$ of the width $\Delta(L)$ 
coincides with the Hurst exponent $H$, as shown here for $H=0.3,0.4,0.5,0.6,0.7,0.8$ }
\label{figscaling}
\end{figure}

In a finite system, the master equation of Eq. \ref{master}
satisfying the detailed balance condition of Eq. \ref{detailedbalance}
will converge exponentially towards the equilibrium Boltzmann distribution.
The characteristic time of this exponential convergence 
is called the equilibrium time $t_{eq}$ 
\begin{eqnarray}
P_t \left({\vec r} \right) - P_{eq}({\vec r})\oppropto _{t \to \infty} e^{- \frac{t}{t_{eq}}} 
\label{defteq}
\end{eqnarray}

Within the strong disorder renormalization procedure described in the 
previous section, this equilibrium time
$t_{eq}$ of a given disordered sample is determined
by the renormalized exit barrier  
\begin{eqnarray}
\Gamma_{eq}= \ln t_{eq}
\label{defgammaeq}
\end{eqnarray}
corresponding to the last decimation process where the two 
largest metastable valleys merge into a surviving valley
corresponding to thermal equilibrium of the whole sample.
We find that the disorder-averaged value $\overline{\Gamma_{eq}}(L)$
and the width $\Delta(L)$ involve the same barrier exponent $\psi$
\begin{eqnarray}
 \overline{\Gamma_{eq}}(L) && \oppropto_{L \to \infty}  L^{\psi} \nonumber \\
\Delta(L)  &&  \oppropto_{L \to \infty} L^{\psi}
\label{psiexp}
\end{eqnarray}
as shown on Fig. \ref{figscaling} (a) for the value $H=0.8$ of the Hurst exponent.
Moreover, this exponent $\psi$ is equal, as expected \cite{Mar83,jpbreview},
 to the Hurst exponent $H$ of the random potential 
\begin{eqnarray}
\psi = H
\label{psieqH}
\end{eqnarray}
We show on Fig.
\ref{figscaling} (b) the log-log plot of the width $\Delta(L)$
for various values $H=0.3,0.4,0.5,0.6,0.7,0.8$ of the Hurst exponent.
Our conclusion is thus that the strong disorder renormalization procedure
confirms the activated nature of the dynamics and
the logarithmic-slow diffusion of Eq. \ref{logslow} in dimension $d=2$.

\begin{figure}[htbp]
 \includegraphics[height=6cm]{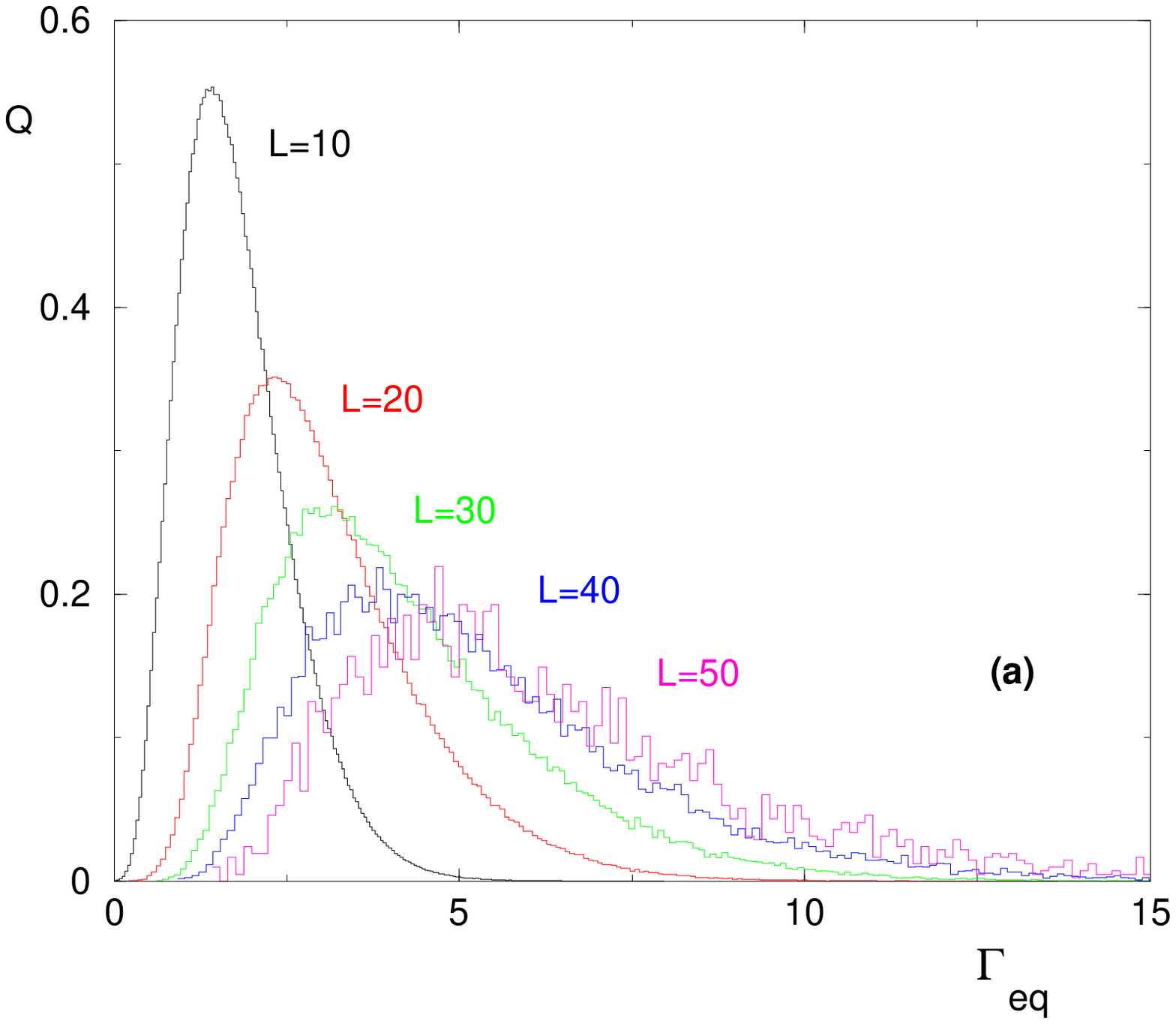}
\hspace{2cm}
 \includegraphics[height=6cm]{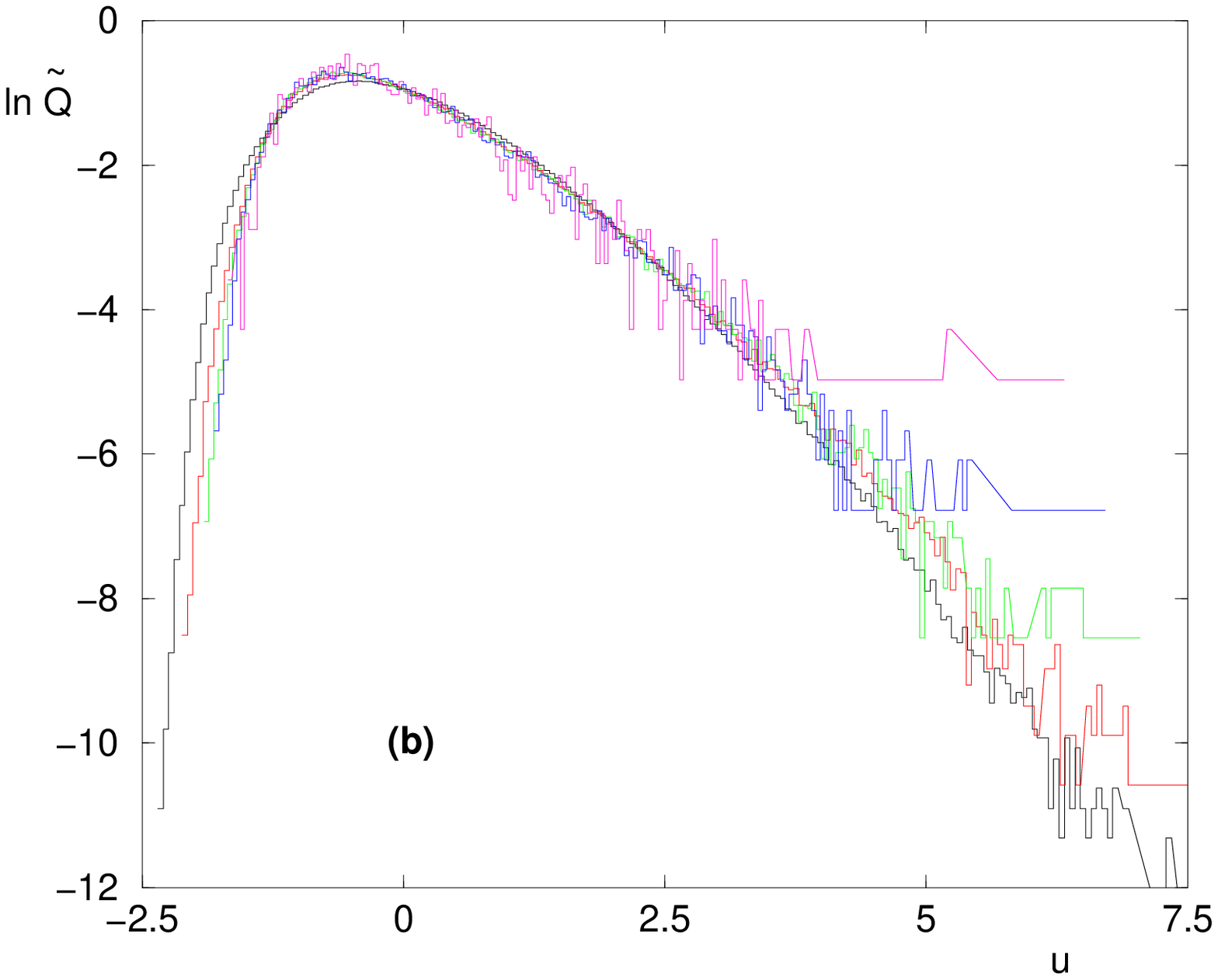}
\vspace{1cm}
\caption{ (Color on line) 
 Statistics of the equilibrium time $t_{eq}$ over the disordered samples of sizes $L^2$ for the case $H=0.8$ :
(a) Probability distribution 
 $Q_{L}(\Gamma_{eq}=\ln t_{eq})$ for $L=10,20,30,40,50$
(b) same data after the rescaling of Eq. \ref{qlteq} and in 
 log-scale to see the tails. }
\label{fighisto}
\end{figure}

We show on Fig. \ref{fighisto} (a) the 
 probability distribution 
 $Q_{L}(\Gamma_{eq}=\ln t_{eq})$
over the disordered samples of size $L^2$ for various sizes $L$, 
for the case $H=0.8$.
 The convergence towards a fixed rescaled distribution 
\begin{eqnarray}
Q_{L}(\Gamma_{eq}) \sim  
  \frac{1}{\Delta(L) } {\tilde Q} 
\left( u \equiv \frac{\Gamma_{eq} - \overline{\Gamma_{eq}}(L) }{\Delta(L) }
 \right)
\label{qlteq}
\end{eqnarray}
 is shown on Fig. \ref{fighisto} (b) in log-scale to see the tails.

\section{ Conclusion }

\label{sec-conclusion}

In this paper, we have shown that the strong disorder renormalization rules
for master equations introduced in \cite{rgmaster} are appropriate
to study random walks in two-dimensional self-affine random potentials
of Hurst exponent $H>0$ : 
we have found an 'Infinite disorder fixed point', 
 where the equilibrium time $t_{eq}$ 
to reach equilibrium for samples of size $L\times L$
scales as $\ln t_{eq}=L^H u $ where $u$ is a random variable of order $O(1)$.
This activated scaling found for the dynamics indicates that
the strong disorder renormalization procedure becomes asymptotically exact
in the limit of large times and large sizes \cite{review}.
Our results confirm that the logarithmic-slow diffusion of Eq. \ref{logslow}
exists not only in dimension $d=1$ where exact results can be obtained
for the Sinai model case $H=1/2$,
but also in higher dimension, as shown here for $d=2$.
These conclusions have been recently checked via independent methods
based on the exact calculation of the biggest relaxation time
 \cite{conjugate} or of some first-passage time \cite{firstpassage}.

\end{document}